\documentclass{ws-mpla}

\begin{document}

\markboth{}
{$\pi K$ interaction  and $CP$ violation in $B\to K\pi^+\pi^-$}

\catchline{}{}{}{}{}

\title{$\pi K$ interaction effects on $CP$ violation\\
  in $B\to K\pi^+\pi^-$ decays
}

\author{ B. LOISEAU}

\address{LPNHE (IN2P3-CNRS-Univ. Paris 6 et 7), Groupe Th\'eorie, Universit\'e Pierre et Marie Curie, 4 place Jussieu, 75252 Paris, France\\
loiseau@lpnhe.in2p3.fr}

\author{B. EL-BENNICH}

\address{Physics Division, Argonne National Laboratory,
Argonne, IL, 60439, USA}

\author{A. FURMAN}

\address{ul. Bronowicka 85/26, 30-091 Krak\'ow, Poland}

\author{R. KAMI\'NSKI, L. LE\'SNIAK}

\address{Division of Theoretical Physics, The Henryk Niewodnicza\'nski Institute of Nuclear Physics,\\
                  Polish Academy of Sciences, 31-342 Krak\'ow, Poland}

\author{B. MOUSSALLAM}

\address{IPN, CNRS/Univ.~Paris-Sud 11, 91406 Orsay Cedex, France}

\maketitle

\pub{Received (Day Month Year)}{Revised (Day Month Year)}

\begin{abstract}
We apply QCD factorization to the quasi two-body $B\to(K\pi)\pi$ decays where the  $(K \pi)$-pair effective mass is limited to 1.8 GeV. 
Our strong interaction phases constrained by theory and $\pi K$ experimental data yield useful information for studies of $CP$ violation.

\end{abstract}

\ccode{PACS Nos.: 13.25Hw,13.75.Lb}

\section{Introduction}	

Three-body charmless non-leptonic $B$ decays provide good opportunities for searches on $CP$ violation in weak interactions and for studies of hadronic physics.
Strong interaction phases are essential for the observation of $CP$ violation and it is important to describe  reliably the strong interaction between the detected hadrons.
Here we study the three-body $B\to K\pi^+\pi^-$ decays for which BaBar and Belle Collaborations have carried out comprehensive Dalitz plot analyses\cite{AubertPRD73,GarmashPRD75}.
For $K\pi$ invariant masses, $m_{K\pi}$, below 1.8 GeV one can observe  the dominance of the vector $K^*(892)$ and of the scalar $K_0^*(1430)$ resonances denoted hereafter as $K^*$ and $K^*_0$.

Even if some phenomenological parameters are necessary, the Quantum Chromodynamics, QCD, factorization approach describes well  $B$ decays into two mesons\cite{LeitnerJPG31}.
This factorization procedure is a leading order approximation in an expansion in inverse powers of the quark $b$ mass $m_b$.
So far no derivation of factorization exists for $B$ decays into three mesons.
However, for $m_{K\pi}\le 1.8$ GeV and in the rest frame of the $B$, the two mesons of the $K\pi$ pair move roughly  in the same direction. 
We refer such processes as $B\to (K\pi)\pi$ and apply factorization to this quasi two-body $B$ decay.
In a preceding work the $B\to(\pi^+\pi^-)K$ decays, the final state of which is dominated by  the $\rho(770)^0$ and $f^0(980)$ resonances, were studied by some of 
us\cite{El-Bennich2006}.

\section{Amplitudes in the QCD factorization approach}

The  $B$-decay amplitudes are  matrix elements of the effective weak Hamiltonian\cite{bene03}
\begin{equation}
\label{Heff1}
H_{eff}={G_F\over\sqrt{2}}\sum_{p=u,c}\lambda_p\, \Big[ C_1 O_1^p + C_2 O_2^p
+\sum_{i=3}^{10} C_i O_i + C_{7\gamma} O_{7\gamma} + C_{8g} O_{8g} \Big] +h.c.\ , 
\end{equation}
where, for strangeness $S=\pm 1$ final states, the $\lambda_p$ are given in terms of the Cabbibo Kobayashi Maskawa, CKM, matrix elements $V_{pp'}$,
$\lambda_u= V_{ub} V^*_{us}$ and $\lambda_c= V_{cb} V^*_{cs}$.
Here,  $G_F$ is the Fermi coupling constant.  The Wilson coefficients $C_i$, associated to the four-quarks operators $O_i$,  depend on the renormalization scale $\mu$.
For the $B^+\to(K^+\pi^-)\ \pi^+$ process, and in the leading order of the factorization method, one has for the penguin operators $O_3$ and $O_4$,
\begin{multline}
\langle \pi^+ ( K^+ \pi^-) \vert C_3 O_3 +C_4 O_4 \vert B^+\rangle=\\
a_4 \langle\pi^+ \vert \bar{b}\gamma^\nu (1-\gamma^5) d\vert B^+\rangle
\langle K^+ \pi^-\vert \bar{s}\gamma_\nu (1-\gamma_5) d\vert 0 \rangle      
\end{multline}
where $a_4= C_4 +C_3/N_c$, $N_c$ being the number of colors. 

The short-distance physics of the weak process $\bar b\to \bar sd\bar d$ enters into this coefficient. 
Its leading order  $O(\alpha_s)$  contribution $a_4 (M_2)$ depends on the nature of the produced meson $M_2$ which, within our quasi two-body approach, we approximate to be the  $K^*_0$ for the $\pi K$-$S$ wave or  the  $K^*$ for the $\pi K$-$P$ wave.
It receives higher order QCD vertex $V_4(M_2)$ and penguin contributions\cite{LeitnerJPG31,bene03}. 
The penguin correction $P_4^{u(c)}(M_2)$ depends not only on  $M_2$ but also on  the intermediate quark, $u$ or $c$, of the studied loop diagram.
Following Refs. 5 and 6 we find, for $\mu=m_b,\ a_4=-0.031$ for both $K^*_0$ and $K^*$, and
$V_4(K_0^*)=-0.001-i0.005,\ V_4(K^*)=-0.002-i0.001,$ $P_4^{u}(K_0^*)=-0.029-i0.019$ and $P_4^{u}(K^*)=-0.004-i0.014$.

The transition $\langle\pi^+\vert\bar b\gamma^\nu d\vert B^+\rangle$ is given in terms
of the scalar and vector $B$ to $\pi$ form factors, 
$F_{0,1}^{B\to\pi}(q^2)$, where the four-momentum  is $q=p_B-p_{\pi^+}=p_K+p_{\pi^-}$.
They contain non-perturbative physics from hadronization of quark currents\cite{LeitnerJPG31,bene03,Cheng06}.
Their $q^2$ variation at low $q^2$ is small and we use 
$F_0^{B^+\to\pi^+}(m_{K_O^*}^2)=0.266$ and $F_1^{B^+\to\pi^+}(m_{K^*}^2)=0.25$.
The matrix element $\langle K\pi\vert\bar q\gamma_\nu s\vert 0\rangle$, given by the strange scalar and vector form factors $f_{0,1}^{K^+\pi^-}(q^2)$, have also long-range physics contributions.
They are calculated\cite{MoussallampiKSandPwave} using analyticity, unitarity, QCD asymptotic counting rules and accurate experimental data on $\pi K$ scattering in the elastic and inelastic domains.

Adding up the next important term  $a_6=C_6+C_5/N_c$ and neglecting other smaller contributions, 
 the $B^+\to(K^+\pi^-)\pi^+$ amplitude is a sum of an $S$- and $P$- wave contributions, 
$\mathcal{M^+} = \mathcal{M}_S^++\mathcal{M}_P^+\ \mathbf{p}_{\pi^-}\cdot\mathbf{p}_{\pi^+}$, with
\begin{multline}
\label{ampliS}
\mathcal{M}_S^+
=\frac{G_F}{\sqrt{2}} (M_B^2-m_\pi^2)\ \frac{m_K^2-m_{\pi}^2}{q^2}\ 
F_0^{B^+\to\pi^+}(q^2) f_0^{K^+\pi^-}(q^2)\\
\times
\bigg(
L_4(K^*_0)-\ \frac{2q^2\ L_6(K^*_0)}{(m_b-m_d)(m_s-m_d)}
\bigg)
\end{multline}
\begin{multline}
\label{ampliP}
\mathcal{M}_P^+
=2\sqrt{2}G_F 
  \ F_1^{B^+\to\pi^+}(q^2)  f_1^{K^+\pi^-}(q^2)
\bigg(L_4(K^*) +\frac{2m_K^*}{m_b}\ \frac{f_V^\perp}{f_V} L_6(K^*)\bigg)
\end{multline}
where $L_i(M_2)=\lambda_u^*\left(a_i^{u}(M_2)+c_i^{u}\right)
+\lambda_c^*\left(a_i^c(M_2)+c_i^{c}\right)$ with $a_{i}^{p}(M_2)=a_{i}(M_2)+V_{i}(M_2)+P_{i}^{p}(M_2)$ and $i=4$ or 6, $p=u$ or $c$.
The term proportional to $f_V^\perp/f_V$ is inferred from the $B\to$ pseudoscalar-vector ($V$) amplitudes\cite{bene03}.
The vertex and penguin corrections are not sufficient to explain the decay rate $B\to K^*\pi$ and we have added four phenomenological complex parameters 
$c_i^{p}$.
These could be interpreted as non perturbative contributions in the penguin diagrams.

For the $\bar B_0\to(\bar K_0\pi^-)_{S,P}\ \pi^+$ amplitude one has the $b\to s\bar u u$ quark transition and, apart from a tree diagram $a_1$ contribution, the derivation is analogous to that of the $B^+\to(K^+\pi^-)_{S,P}\ \pi^+$.
The $B^-$ and $B_0$ amplitudes are obtained from the $B^+$ and $\bar B_0$ ones with the replacements $\lambda_u^*$ and $\lambda_c^*$ into $\lambda_u$ and $\lambda_c$.

\begin{figure}[h]
\begin{center}
\includegraphics[scale=0.35] {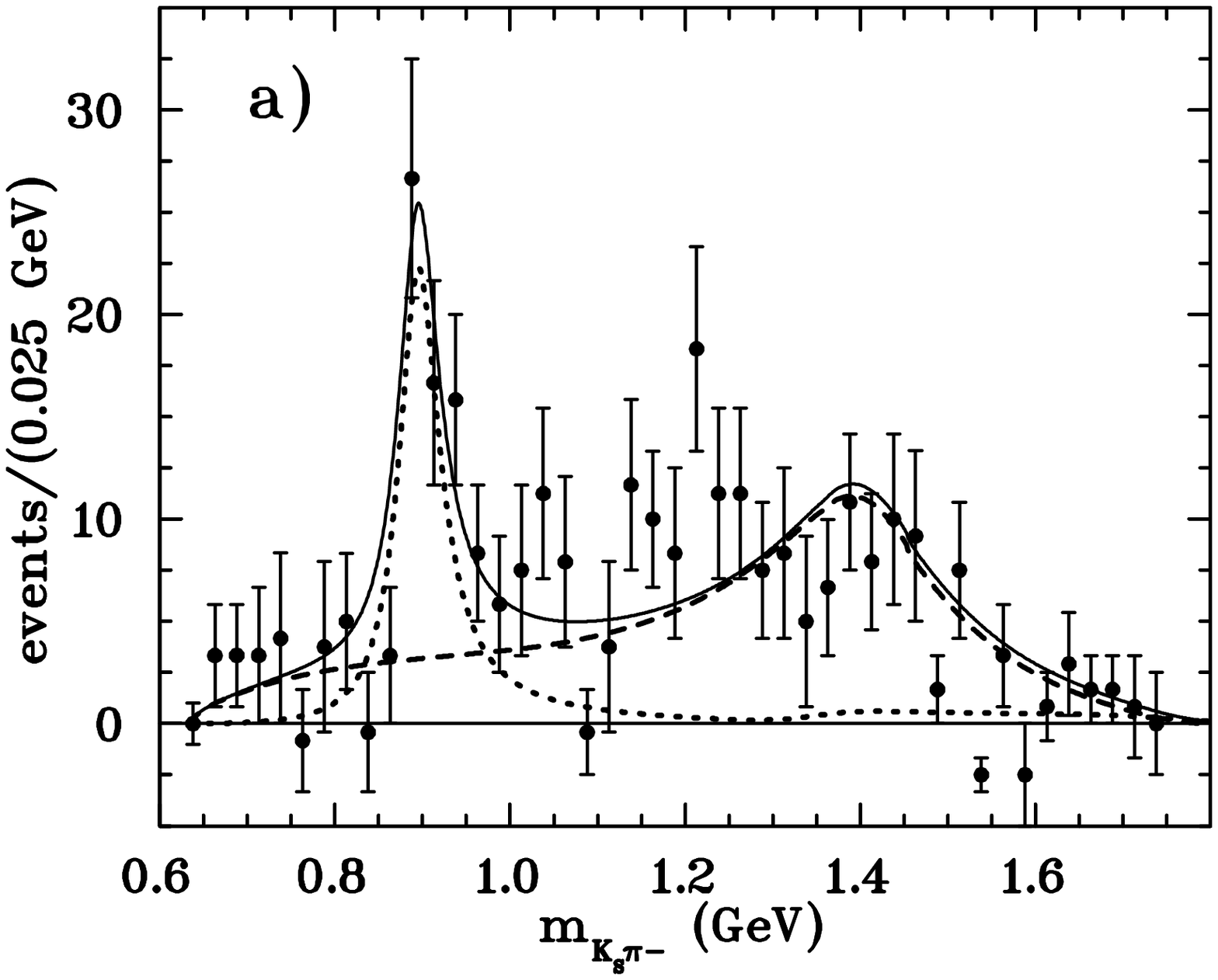}\hspace{1cm}
\includegraphics[scale=0.35] {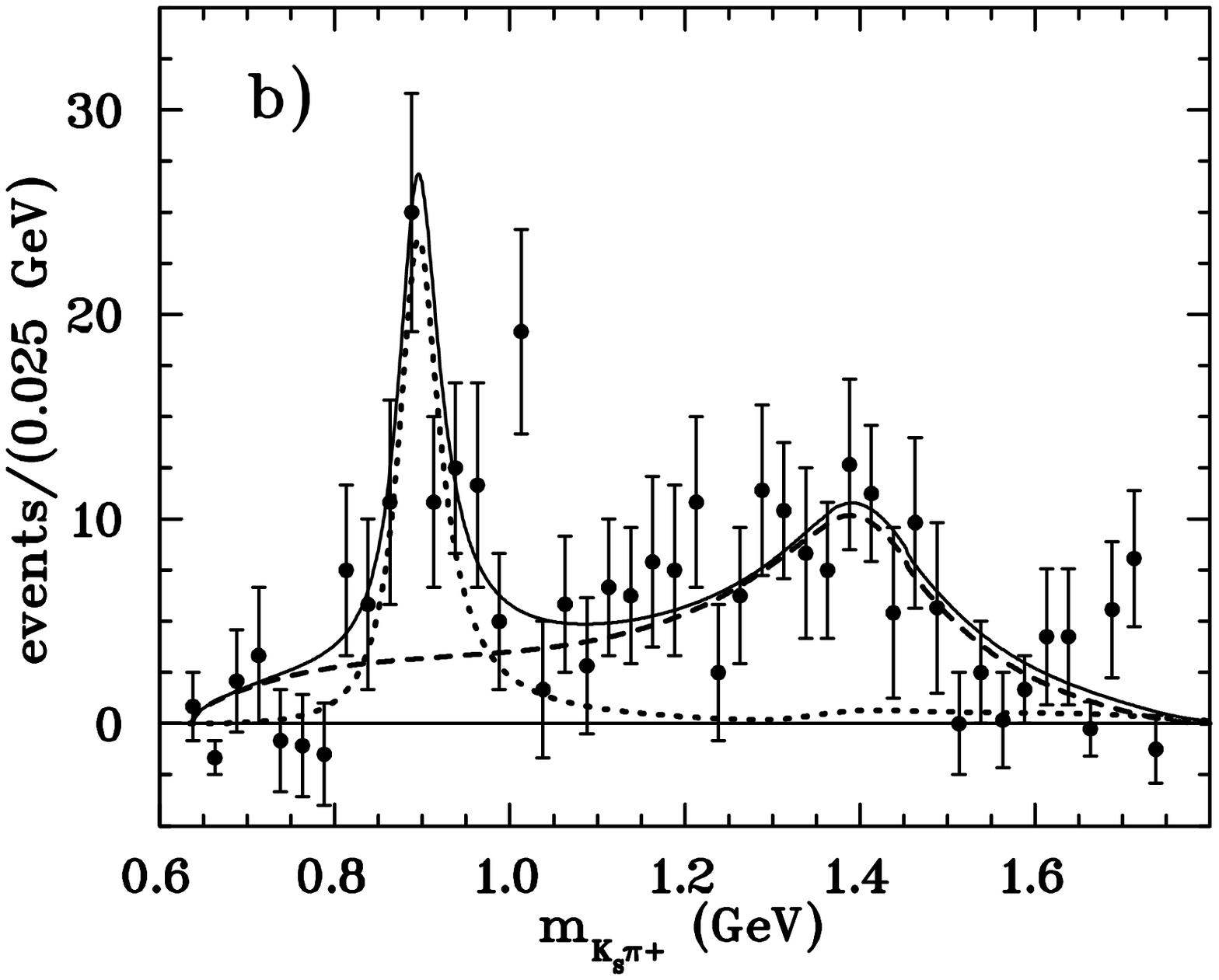}
\caption{Some results from a preliminary fit to the $B\to K\pi^+\pi^-$ data. Comparison with the experimental $K\pi$ effective mass distributions for: a) $\bar B^0\to\bar K^0\pi^+\pi^-$ and b) $B^0\to K^0\pi^+\pi^-$. The ($K\pi$) $P$-wave contribution is drawn as dotted lines, that of the $S$ wave as dashed lines and their sum as solid lines. The data are from the Belle Collaboration\protect\cite{GarmashPRD75}.}
\label{fig:gdh}
\end{center}
\end{figure}

\section{Results and conclusions}
We have performed a preliminary fit on the four $c_{i}^{p}$ ($i=4$ or 6, $p=u$ or $c$). A more complete analysis will appear soon\cite{El-Bennich2008}.
Our model reproduces well the $m_{K\pi}$ distributions in both $K^*(892)$ and $K_0^*(1430)$ regions (see Fig. 1).
The $S$-wave contribution enhancement below 1 GeV, related to the $K^*(800)$ (or $ \kappa$) resonance, is well described.
The resulting $m_{\pi K}$ helicity-angle distributions compare well with experiment and show interesting $S$- and $P$-wave interference effects.
The $B\to K^*(892)\pi$ branching ratios are close to the experimental values, but those for $B\to K_0^*(1430)\pi$, compatible with a new BaBar analysis, are smaller than the experimental numbers from Belle.
The $CP$ asymmetries are comparable to the experimental values.

In conclusion the use of the strange form factors\cite{MoussallampiKSandPwave}, constrained by experiments other than those from $B$ decays and by theory, is an alternative to the isobar model often used in the experimental analysis of the Dalitz plots.
Furthermore it allows us to calculate, using the complex pole definition of a resonance, the branching ratios and the decay constants of the $K^*(892)$ and $K_0^*(1430)$.
We have looked at the effect of the next to leading order in $\alpha_s$, vertex and penguin correction in the framework of the QCD factorization, it would be  interesting to include also hard spectator scattering and annihilation\cite{LeitnerJPG31,bene03,Cheng06}.
Nevertheless, our strong interaction phases constrained by theory and $\pi K$ experimental data should provide useful information for studies of $CP$ violation.

\section*{Acknowledgements}

We acknowledge helpful comments from J-P. Dedonder and O. Leitner.
This work was supported by the agreements between IN2P3 and Polish Laboratories (collaboration N$^\circ$ 08-127), betwen PAN and CNRS (collaboration N$^\circ$ 19481) and by the Department of Energy, Office of Nuclear Physics, contract n$^0$ DE-AC02-06CH11357.



\begin{thebibliography}{99}

\bibitem {AubertPRD73} 
 B. Aubert, \textsl{et al.}, BaBar Collaboration,  Phys. Rev. D {\bf 73}, 031101(R) (2006), 
 {\it Measurements of neutral $B$ decay branching fractions to  $K^0_S \pi^+ \pi^-$
final states and the charge asymmetry of $B^{0} \to K^{*+} \pi^-$}.

\bibitem {GarmashPRD75}
A. Garmash,  \textsl{et al.}, Belle Collaboration,  Phys. Rev. D {\bf 75}, 012006 (2007), 
{\it  Dalitz analysis of the three-body charmless $B^{0} \to K^0 \pi^+ \pi^-$ decay.}

\bibitem{LeitnerJPG31} 
M. Beneke, Nucl. Phys. B (Proc. Suppl.) \textbf{170}, 57 (2007),  \textit{Hadronic B decays};
O. Leitner, X-H. Guo, A.W. Thomas,
J. Phys. G: Nucl. Part. Phys. \textbf{31}, 199 (2005),
\textit{Direct $CP$ violation, branching ratios and form factors $B\to\pi,\ B\to K$ in $B$ decays.}

\bibitem{El-Bennich2006}
   B. El-Bennich, A. Furman, R. Kami\'nski, L.~Le\'sniak and B.~Loiseau,
   Phys. Rev. D \textbf{74}, 114009 (2006),
   \textit{Interference between $f_0(980)$ and $\rho(770)^0$ resonances in $B\to\pi^+\pi^-K$ decays}.

 \bibitem{bene03}
  M.~Beneke and M.~Neubert,
  Nucl. Phys. B \textbf{675}, 333 (2003), 
  \textit{QCD factorization for $B\to PP$ and $B\to PV$ decays}.
  
  \bibitem{Cheng06}
  H-Y. Cheng, C-K. Chua and K-C Yang,
Phys. Rev. D\textbf{73}, 014017 (2006),
\textit{Charmless hadronic B decays involving scalar mesons: Implications to the nature of light scalar mesons.} 
  
  \bibitem{MoussallampiKSandPwave} 
B. Moussallam,  Eur. Phys. J. \textbf{C53}, 401 (2008), 
 \textit{Analyticity constraints on the strangeness changing vector current and applications to $\tau\to K\pi \nu_\tau$, $\tau\to K\pi\pi \nu_\tau$.} 
 
 \bibitem{El-Bennich2008}
   B. El-Bennich, A. Furman, R. Kami\'nski, L.~Le\'sniak, B.~Loiseau, B.~Moussallam, \textit{ $CP$ violation and pion-kaon interactions  in $B\to K\pi^+\pi^-$}decays, work in preparation.

 
 
\end{thebibliography}
\end{document}